\documentclass[preprint]{aastex}
\usepackage{emulateapj5}
\usepackage{apjfonts}
\usepackage{natbib}

\input psfig.tex

\def\hst{{\it HST}}
\def\etal{\emph{et al.}\ }

\def\kms{km s$^{-1}$}
\def\apj{ApJ}
\def\aj{AJ}
\def\mnras{MNRAS}
\def\apjs{ApJS}

\def\aj{{AJ}}

\def\apj{{ApJ}}
\def\apjs{{ApJS}}
\def\asec{$^{\prime\prime}$}

\def\deg{$^{\circ}$}

\def\etal{{et al. }}

\def\mnras{{MNRAS}}
\def\nat{{Nature}}

\def\percm2{cm$^{-2}$}

\def\lax{{$\mathrel{\hbox{\rlap{\hbox{\lower4pt\hbox{$\sim$}}}\hbox{$<$}}}$}}
\def\gax{{$\mathrel{\hbox{\rlap{\hbox{\lower4pt\hbox{$\sim$}}}\hbox{$>$}}}$}}

\slugcomment{Submitted to {\it ApJ}\ on July 20, 2001; accepted 
on October 18, 2001.}
\lefthead{RAVINDRANATH, HO, \& FILIPPENKO}
\righthead{NUCLEAR CUSPS AND BINARY BLACK HOLES}

\begin{document}

\title{Nuclear Cusps and Cores in Early-type Galaxies As Relics of 
Binary Black Hole Mergers} 

\author{
Swara Ravindranath\altaffilmark{1,2},
Luis C. Ho\altaffilmark{1}, and 
Alexei V. Filippenko\altaffilmark{2} \\ 
}

\altaffiltext{1}{The Observatories of the Carnegie Institution of Washington, 
813 Santa Barbara St., Pasadena, CA 91101-1292.}

\altaffiltext{2}{Department of Astronomy, University of California, 
Berkeley, CA 94720-3411.}

\setcounter{footnote}{3}

\begin{abstract}
We present an analysis of the central cusp slopes and core parameters of 
early-type galaxies using a large database of surface brightness profiles 
obtained from {\it Hubble Space Telescope}\ observations.  We examine 
the relation between the central cusp slopes, core parameters, and black hole 
masses in early-type galaxies, in light of two models that attempt to explain 
the formation of cores and density cusps via the dynamical influence of black 
holes. Contrary to the expectations from adiabatic-growth models, we find that 
the cusp slopes do not steepen with increasing black hole mass fraction. 
Moreover, a comparison of kinematic black hole mass measurements with the 
masses predicted by the adiabatic models shows that they overpredict the
masses by a factor of $\sim$3. Simulations involving binary black hole mergers
predict that both the size of the core and the central mass deficit 
correlate with the final black hole mass. These relations are qualitatively 
supported by the present data. 
\end{abstract}


\section{Introduction}                   

High-resolution images obtained using ground-based and {\it Hubble Space 
Telescope (HST)}\ observations have shown that density cusps are ubiquitous in 
the centers of early-type galaxies and true constant-density cores are 
extremely rare (Kormendy 1985; Lauer 1985; Crane et al. 1993; Lauer et al. 
1995). The central surface brightness profiles can be parameterized 
by a single power law in galaxies where no core is resolved (power-law 
galaxies), while others require a double power law such that the slope within 
the resolved core is much shallower than the outer slope (core galaxies). 
According to the widely adopted criteria of Lauer et al. (1995) and Faber et 
al. (1997), the central surface brightness cusps can be described by 
$I(r)\,\propto \,r^{-\gamma}$, with $\gamma$ $\leq$ 0.3 for core galaxies and 
$\gamma$ $\geq$ 0.5 for power-law galaxies.

Interestingly, this classification based on the cusp slope reflects a distinct 
dichotomy in the global photometric and kinematic properties among early-type 
galaxies. Core galaxies are luminous ($M_{V}$ \lax\ $-21$ mag) and have high 
velocity dispersions, compared to power-law galaxies which have lower 
luminosities and lower velocity dispersions. Core galaxies have boxy or pure 
elliptical isophotes and show slow rotation, while power-law galaxies tend to 
have disky isophotes and are fast rotators (Carollo et al. 1997; Faber et al. 
1997; Ravindranath et al. 2001b; Rest et al. 2001).  
Differences between the two 
classes can also be seen in their X-ray properties (Pellegrini 1999). Thus, 
it appears that the origin of the inner cusps is closely linked to the 
formation and subsequent evolution of these galaxies.

Several models with and without central, massive black holes (BHs) have been 
proposed to explain the formation of cusps and the observed core properties. 
In models without BHs the cusp slope reflects the amount of dissipation that 
occurred during the formation process. Steep density cusps seen in 
low-luminosity elliptical galaxies can result from dissipative, gas-rich 
mergers (Faber et al. 1997). Numerical simulations of gas-rich mergers have 
shown that angular momentum transfer can lead to gas inflows, trigger 
starbursts, and result in the formation of dense cores (e.g., Barnes \& 
Hernquist 
1991). Moreover, even a small amount of gas can cause the isophotes to become 
disky and induce the higher rotation velocities seen in power-law galaxies 
(Barnes \& Hernquist 1996; Kormendy \& Bender 1996; Faber et al. 1997).  An 
analogous explanation for the weak cusps seen in luminous ellipticals involves 
dissipationless, gas-poor mergers (Bender, Burstein, \& Faber 1992). Although 
such models succeed in producing the weaker cusps seen in these objects, the 
core sizes tend to decrease or remain unchanged in the merger remnant 
(Farouki, Shapiro, \& Duncan 1983; Barnes 1992). In fact, with the increased 
merger events that would be needed to form giant ellipticals, the cores become 
denser and more compact, in clear conflict with the empirical trend that giant 
ellipticals tend to have large-sized, diffuse cores. Even in the case of 
accretion of a low-mass secondary galaxy by a massive primary galaxy, the 
dense core of the secondary survives the merger event, thereby resulting in a 
remnant that has high central density (Holley-Bockelmann \& Richstone 1999, 
2000; Merritt \& Cruz 2001). Thus, models that aim to account for the 
formation of central density cusps without invoking BHs invariably produce 
cusps that are steeper than observed and fail to reproduce the scaling 
relations between galaxy luminosity and core size.
 
On the other hand, numerical simulations that include the effects of central 
BHs have proved to be more promising in reproducing the low-density cores 
seen in luminous ellipticals (Ebisuzaki, Makino, \& Okumura 1991; Makino \& 
Ebisuzaki 1996; Nakano \& Makino 1999; Milosavljevi\'c \& Merritt 2001). 
Ebisuzaki et al. (1991) showed that the observed correlation between core size 
and total bulge luminosity can result from mergers of equal-mass galaxies 
containing central BHs. More recent studies have shown that the formation and 
subsequent orbital decay of a BH binary can destroy the existing steep cusps 
of the merging galaxies via the interaction of the binary with surrounding 
stars. During the initial stages of merging, the BHs sink to the center under 
the influence of dynamical friction from stars and form a binary. The potential
energy released as the binary orbit shrinks can eject stars in its vicinity 
with high velocities, thereby weakening the initial density cusp and 
expanding the core (Quinlan \& Hernquist 1997; Nakano \& Makino 1999; 
Milosavljevi\'c \& Merritt 2001). The merger models predict that the core 
size of the merger remnant would show an almost linear dependence with BH 
mass (Nakano \& Makino 1999). During each merger event the scouring 
action by the BH binary would eject a total mass of the order of the combined 
mass of the BHs (Quinlan \& Hernquist 1997; Milosavljevi\'c \& Merritt 2001).
In addition to the BH mass, the overall mass deficit depends on the number of 
merger events, and is larger for the more luminous galaxies. Thus, a 
steeper-than-linear relation is predicted between the ejected mass and the 
final BH mass (Milosavljevi\'c \& Merritt 2001).  

An alternative model interprets the central cusps in elliptical galaxies as 
the result of dynamical effects induced by the adiabatic growth of a central 
BH in an initially isothermal core (e.g., Young et al. 1978; Lauer et al. 
1992; Crane et al. 1993; van~der~Marel 1999a).  The growth of BHs in galaxy 
centers can occur by accretion of gas from stellar mass loss or other sources. 
Under the adiabatic assumption this growth is slow compared to the dynamical 
timescale, such that the phase-space distribution function of the stars 
remains unchanged. Adiabatic growth in an isothermal core induces a weak 
density cusp of the form $\rho\,\propto \,r^{-3/2}$, or slightly steeper for 
the case of a nonisothermal core (Young 1980; Quinlan, Hernquist, \& 
Sigurdsson 1995).  According to these models, the slope of the cusp is 
determined by the ratio of the BH mass to the mass of the initial core. A 
steeper cusp would result from a larger fractional BH mass. This is in 
contrast to the expectations of the merger models, where a more massive BH 
would cause the cusp to weaken.  The cusp slope and the cusp scale length of 
the surface brightness profiles derived from adiabatic-growth models is found 
to correlate well with the BH mass (Cipollina \& Bertin 1994). Van~der~Marel 
(1999a) has shown that the adiabatic models can reproduce the range of 
inner-cusp slopes and core radii seen in the \hst\ images and can be used to 
predict BH masses.

In this paper, we use a large database of observed central parameters of 
early-type galaxies to test the viability of the adiabatic BH growth and 
binary BH merger models.  High-resolution surface brightness profiles are 
currently available for many early-type galaxies.  The discovery of a tight 
correlation between BH mass and bulge velocity dispersion (Gebhardt et al. 
2000; Ferrarese \& Merritt 2000) allows us to estimate BH masses accurately 
and  efficiently.  These recent developments provide an opportunity to test 
the models.  Section 2 describes the sample and the sources from which the 
data were compiled.  We present the observed correlations in \S~3 and discuss 
their implications in \S~4.   A preliminary version of this work was 
presented in Ravindranath, Ho, \& Filippenko (2001a).

\section{The Sample}        

The sample chosen for the present study comprises early-type galaxies that 
have been observed using \hst\ and for which their central surface brightness 
profiles (within a radius of $\sim$10\asec) have been parameterized in the 
literature using the ``Nuker'' function (Lauer et al. 1995).  This function 
has the form

\begin{equation}
I(r)=2^{(\beta-\gamma)/\alpha}I_{b}\left(\frac{r}{r_{b}}\right)^{-\gamma}
\left[1+\left(\frac{r} {r_{b}}\right)^{\alpha}\right]^{(\gamma-\beta)/\alpha},
\end{equation}

\vspace*{0.2cm}

\noindent
where $\beta$ is the asymptotic slope as $r \rightarrow \infty$, $\gamma$
is the asymptotic slope as $r \rightarrow 0$, $r_{b}$ is the break radius at 
which the outer slope $\beta$ changes to the inner slope $\gamma$, $\alpha$ 
controls the sharpness of the transition between the inner and outer slopes, 
and $I_{b}$ is the surface brightness at $r_{b}$.  

The sample is derived mainly from three sources.  (1) Faber et al. (1997) give 
Nuker parameters for 58 nearby early-type galaxies that were observed using 
WFPC in the $V$ band. These include galaxies in the Local Group, ellipticals 
and S0s in the Virgo cluster, and the bulges of a few nearby early-type spiral 
galaxies. (2) Rest et al. (2001) obtained $R$-band images of 67 early-type 
galaxies using WFPC2 and fitted Nuker functions to the light profiles of 58 
of these.  Their sample comprises all early-type galaxies from the Lyon/Meudon 
Extragalactic Database with radial velocities less than 3400 \kms, absolute 
$V$-band magnitudes less than --18.5, and Galactic latitudes greater than 
20\deg.  (3) Ravindranath et al. (2001b) used the Nuker function to perform
two-dimensional fits of the central regions of 33 early-type galaxies 
observed with NICMOS in the $H$ band. The galaxies were chosen from the 
Palomar survey of nearby galaxies, a ground-based optical spectroscopic study
of a nearly complete sample of 486 galaxies with $B_{T}$ $\leq$ 12.5 mag and 
$\delta\,>$ 0\deg\ (Filippenko \& Sargent 1985; Ho, Filippenko \& Sargent 1995).

From these sources, we compiled a database consisting of galaxies for which 
the classification of the central surface brightness profile type is 
unambiguous. Only galaxies with inner slope $\gamma$ $\leq$ 0.3, break radius 
$r_b\,>$ 0\farcs2, and $\alpha\,>$ 0.8 were included as core galaxies, to 
ensure that the core region is well resolved.  Objects with $\gamma\,\geq$ 0.5 
constitute power-law galaxies in the sample. We did not include in the 
present analysis the few objects that have intermediate slopes 
(0.3 $<\,\gamma\,<$ 0.5).  We also excluded galaxies lacking stellar velocity 
dispersion measurements, which we later use to estimate BH masses (\S~4).  The 
final database consists of 88 sources (38 core galaxies and 50 power-law 
galaxies), the properties of which are given in Tables 1A and 1B, respectively. 
 
Since the above data were taken in different bands, it is important to verify 
whether the central parameters vary with wavelength, as would be expected if 
there are significant differences in the stellar population or dust extinction. 
Ravindranath et al. (2001b) compared the Nuker parameters derived from optical 
and near-infrared images. Even though the outer slope $\beta$ can be 
significantly different between the two bands (a consequence of radial color 
gradients), the inner slope $\gamma$ and the break radius $r_b$ show good 
overall agreement. Another concern when using data from various sources is to 
ensure that physical parameters are derived using the same distance scale.  The 
apparent quantities from the literature were converted to absolute quantities 
using distances from Tonry et al. (2001), when available, which are based on 
measurements of $I$-band surface brightness fluctuations.  The distances for 
the remaining galaxies were calculated from the radial velocities given in the 
Third Reference Catalogue of Bright Galaxies (de~Vaucouleurs et al. 1991) 
using a Hubble constant of $H_0$ = 75 \kms\ Mpc$^{-1}$. 


\section{Comparison of Data with Model Predictions} 

Both the adiabatic-growth model and the binary BH model attempt to explain the 
formation of cores and weak cusps in elliptical galaxies based on the 
dynamical effects of BHs. We compare the relation between the observed 
quantities (core radius, inner-cusp slope, and BH mass) with the relations 
predicted from these models. The BH masses required for the analysis presented 
here are derived using the empirical correlation between BH mass and bulge 
stellar velocity dispersion (Gebhardt et al. 2000; Ferrarese \& Merritt 
2000).  Gebhardt et al. (2000) use $\sigma = \sigma_e$, the projected, 
luminosity-weighted velocity dispersion measured within the effective radius 
$r_e$. 
%
%
\begin{figure*}[t]
\centerline{\psfig{file=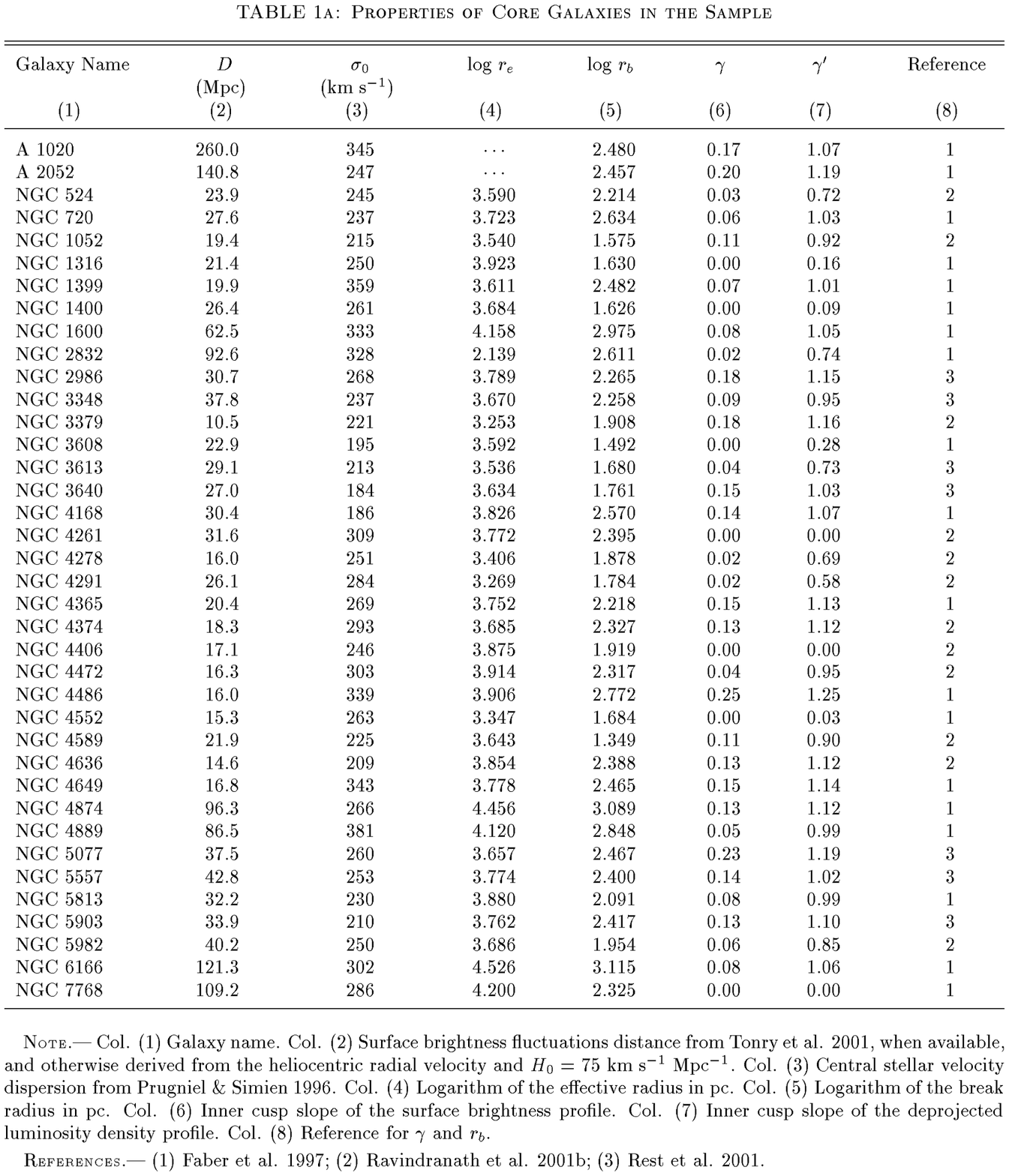,width=17.5cm,angle=0}}
\end{figure*}



\begin{figure*}
\centerline{\psfig{file=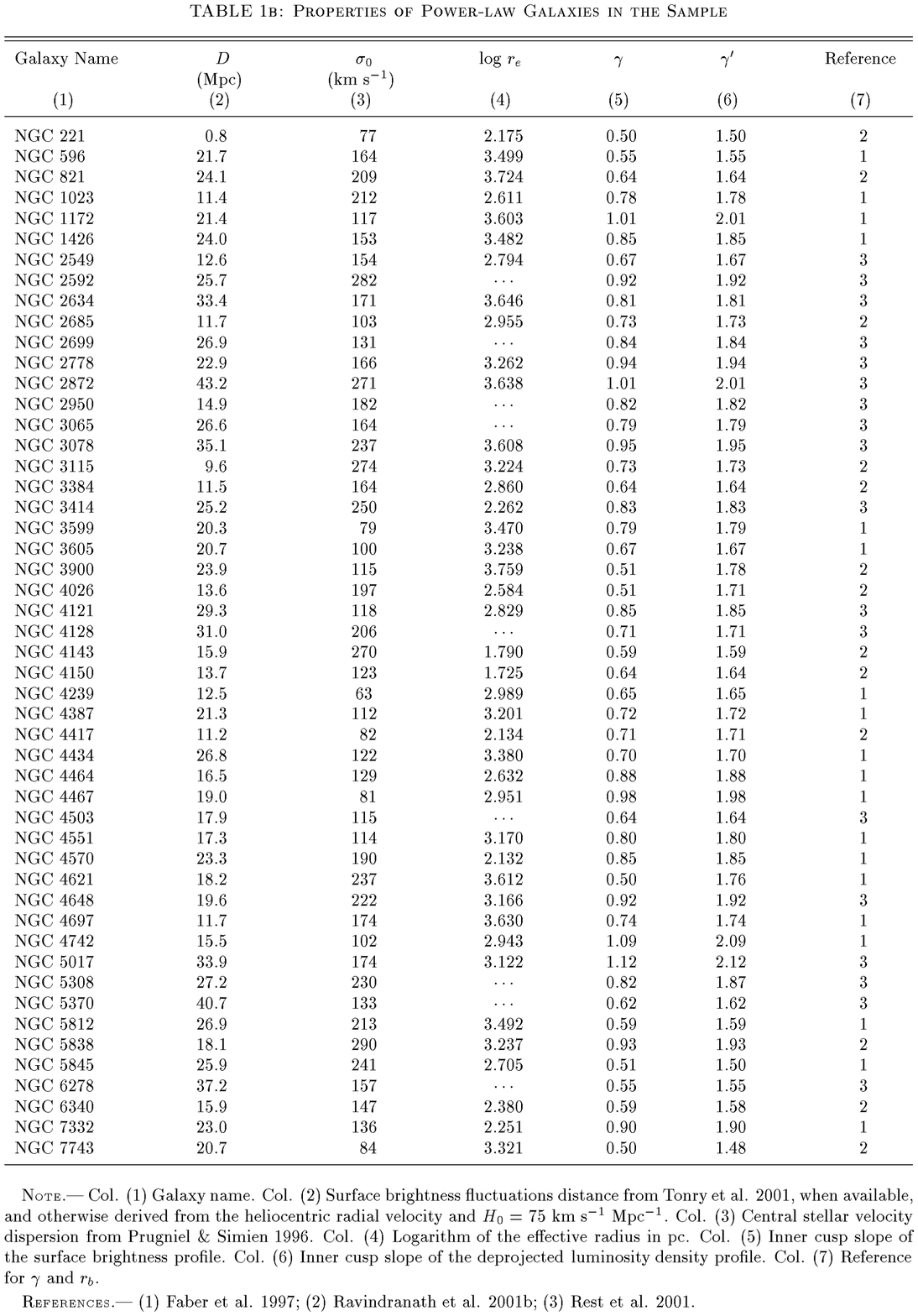,width=17.5cm,angle=0}}
\end{figure*}


\noindent
For our sample we do not have uniform measurements of
$\sigma_e$, we use instead $\sigma = \sigma_0$, the central velocity
dispersion, since in general $\sigma_e\,\approx\, \sigma_0$ within a scatter of
$\sim$10\% (Gebhardt et al. 2000).  The central velocity dispersions were taken 
from the compilation by Prugniel \& Simien (1996).

\vspace*{0.2cm}
\subsection{Cusp Slopes and Black Hole Masses}

The adiabatic-growth model predicts that more massive BHs (relative to galaxy
core mass) produce steeper density cusps. This is well illustrated when the 
inner-cusp slope ($\gamma$ as parameterized in the Nuker function) is plotted 
against the dimensionless BH mass $\mu\,\equiv\,M_{\rm BH}/M_{\rm core}$ 
(Cipollina \& Bertin 1994; van~der~Marel 1999a), where $M_{\rm core}$
is the mass of the initial isothermal core.  We do not have an independent 
estimate of $M_{\rm core}$, since the observed core properties, by assumption, 
have been altered by the BH. However, if the initial state of the galaxy can be 
approximated by an isothermal sphere (as is done in the adiabatic models), the 
initial core radius would be a fixed fraction of the initial half-mass radius. 
Since the 
%
\vskip 0.3cm
\psfig{file=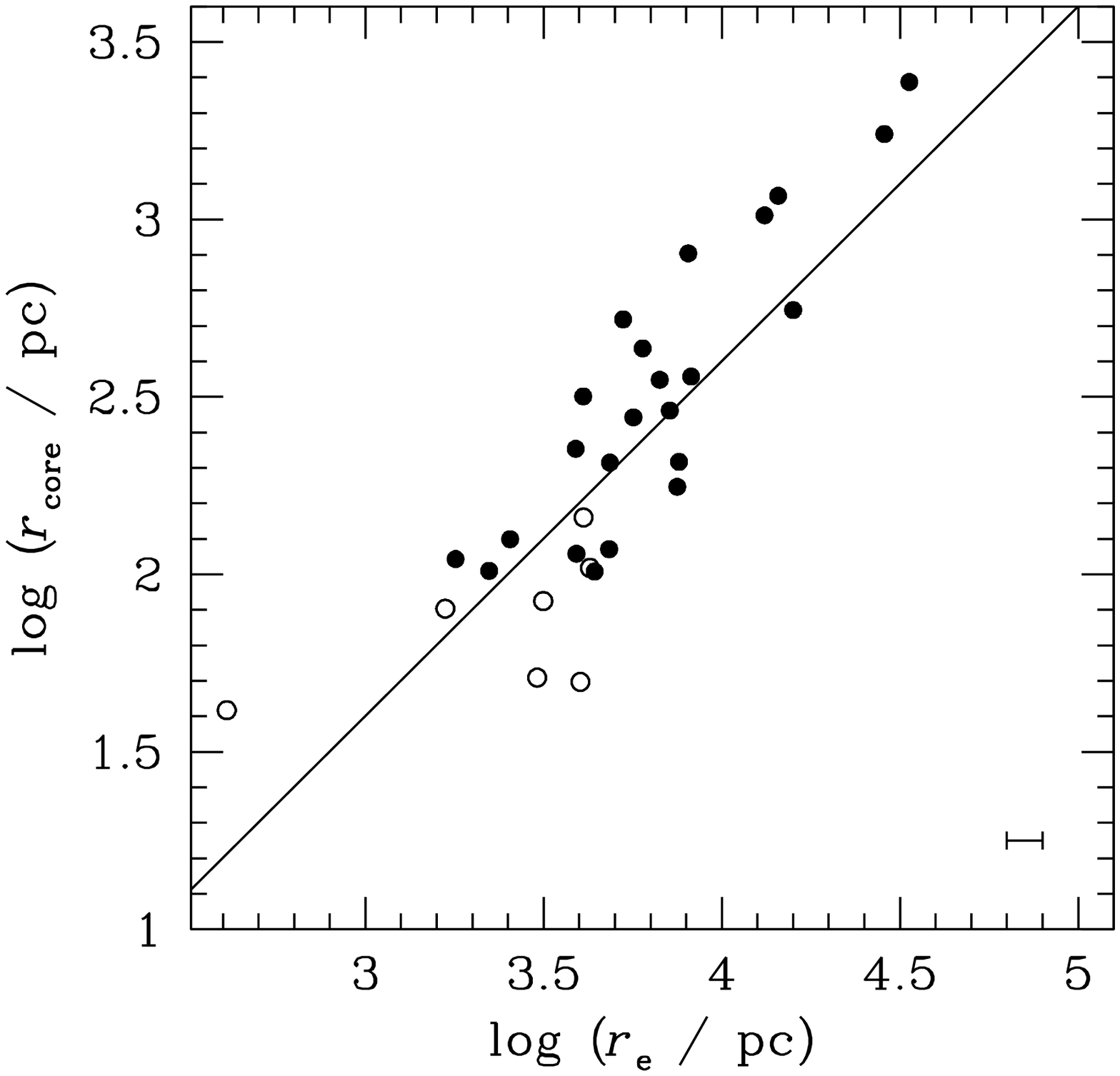,width=8.5cm,angle=0}
\figcaption[f1.ps]{
The radius of the initial isothermal core for the best-fitting
adiabatic-growth model versus the effective radius. Core galaxies are shown
as solid points and power-law galaxies as open circles. The solid line has
a slope of unity and corresponds to {\it r}$_{\rm core}$ = 0.04 {\it r}$_{e}$.
Only galaxies common to this study and that of van~der~Marel (1999a) are
shown.  A representative error bar for {\it r}$_{e}$ is given in the
lower-right corner of the plot.
\label{fig1}}
\vskip 0.3cm


\noindent
adiabatic growth of the BH should not significantly alter the mass 
distribution at large radii, it is safe to assume that the half-mass radius, 
or a closely related measure such as the effective radius, remains 
unaffected.  In Figure~1, we plot the core radius from the best-fitting 
adiabatic model of van~der~Marel (1999a) versus the effective radius for all 
galaxies common to both studies.  The radius of the isothermal core indeed is 
well correlated with $r_{e}$ and can be approximated as 
$r_{\rm core}$ $\approx$ 0.04 $r_{e}$; the mean scatter about this relation is 
0.23~dex. Thus, analogous to $\mu$, we define the dimensionless mass 
$\mu^{\prime}\,\equiv\,M_{\rm BH}/M_{e}$, where $M_{e}$ is the mass within 
$r_e$ calculated using $M_{e} = 3 \sigma^2 r_e / G$.  The effective radii were 
taken from Faber et al. (1989), Bender et al. (1992), and Baggett, Baggett, 
\& Anderson (1998).

A plot of the observed inner-cusp slope $\gamma$ versus $\mu^{\prime}$ 
(Fig.~2) does not appear to be consistent with the expectation from the 
adiabatic-growth model.  We have overplotted the theoretical relation between 
$\gamma$ and $\mu$, as given  by Cipollina \& Bertin (1994), but shifted 
arbitrarily in the horizontal direction to allow for a constant scaling 
between $\mu$ and $\mu^{\prime}$ (i.e., between $M_{\rm core}$ and $M_{e}$).
The predicted line matches the data poorly.  The largest discrepancy lies with 
the power-law galaxies, which, because of their steep cusps, are required to 
have values of $\mu^{\prime}$ up to 1--2 orders of magnitude larger than 
observed.  Except for a larger spread, most power-law galaxies, in fact, are 
characterized by values of $\mu^{\prime}$ within the fairly restricted range 
found in core galaxies ($\log \mu^{\prime}\,\approx\,-2.9\pm0.5$).  In other 
words, the BH mass constitutes a roughly constant fraction of the central 
stellar mass within $r_e$, albeit with substantial scatter.  This is just the 
familiar linear correlation between BH mass and bulge mass (Kormendy \& 
Richstone 1995; Magorrian et al. 1998; van~der~Marel 1999b; Ho 1999; 
Kormendy \& Gebhardt 2001).  The distribution of points for the core galaxies 
might be marginally consistent with the model, but the scatter is large.

Since $M_{e}$ is distance dependent, this test may be affected, at least in 
part, by uncertainties in the distances used.  The parameter $\gamma$ itself 
is also vulnerable to distance effects because the 
%
\vskip 0.3cm

\psfig{file=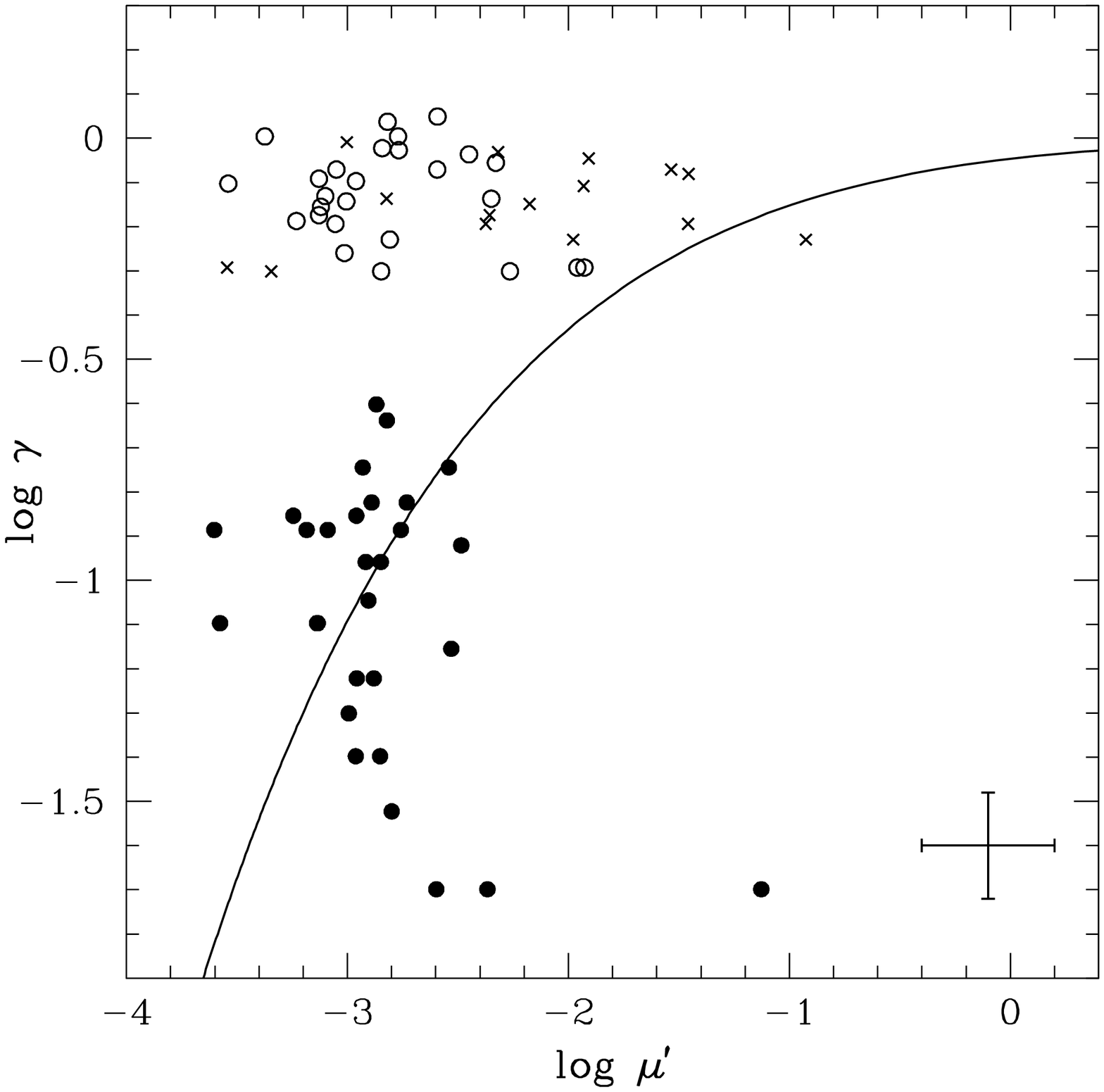,width=8.5cm,angle=0}
\figcaption[f2.ps]{
The inner-cusp slope $\gamma$ versus the dimensionless BH mass defined as
$\mu^{\prime}\,\equiv\,M_{\rm BH}/M_{e}$, where $M_{e}$ is the total mass
within the effective radius. Core galaxies are shown as solid points and
power-law galaxies as open circles. Crosses denote galaxies for which the
effective radii were derived from bulge-disk decomposition (from Baggett et
al. 1998).  Representative error bars are given in the
lower-right corner of the plot.  The solid line gives the theoretical
prediction from Cipollina \& Bertin (1994) for the relation between $\gamma$
and $\mu\,\equiv\,M_{\rm BH}/M_{\rm core}$, shifted arbitrarily along the
horizontal axis to allow for a constant scaling between $\mu$ and
$\mu^{\prime}$.
\label{fig2}}
\vskip 0.3cm


\noindent
cores of distant galaxies 
may not be sufficiently well resolved to yield the asymptotic inner slope of 
the Nuker function.   This latter effect, however, cannot be strong for the 
core galaxies in our sample because we have explicitly chosen objects with 
well-resolved break radii (see \S~2).  We do not believe that distance errors 
could have erased an intrinsic trend in Figure~2.  The nominal error bars 
shown on the lower-right corner of the plot are significantly smaller than the 
observed spread in the data points.  

Figure~3{\it a}\ compares the BH masses inferred from adiabatic models 
by van~der~Marel (1999a) with independent, direct measurements obtained from 
stellar and gaseous kinematics (as summarized in Kormendy \& Gebhardt 2001); a 
similar comparison is made in Figure~3{\it b}\ for a larger sample with  
independent masses derived using the $M_{\rm BH}-\sigma$ relation of Gebhardt 
et al. (2000).  The adiabatic models tend to yield masses that are 
systematically larger, on average by $\sim$0.5 dex; the offset is slightly 
larger for power-law galaxies than core galaxies.

\subsection{Core Properties and Black Hole Binaries}

Simulations of merging galaxies with central BHs produce merger remnants 
whose core properties correlate well with the final BH mass (Ebisuzaki et al. 
1991; Nakano \& Makino 1999; Milosavljevi\'c \& Merritt 2001). Nakano \& Makino 
(1999) studied the dynamical reaction of the central regions of a galaxy, 
modeled as an isothermal core, to the infall of a massive BH. They found that 
larger BHs tend to create larger and more massive cores. Figure~4{\it a}\ 
shows the relation between the observed core size --- here represented by the 
break radius $r_b$ --- and the BH mass.  Albeit with substantial scatter, 
the observed trend qualitatively agrees with the simulations in that more 
massive BHs produce larger cores.  The correlation revealed by our sample is 
highly statistically significant: the Spearman's $\rho$ 

\vskip 0.3cm

\begin{figure*}[t]
\figurenum{3}
\centerline{\psfig{file=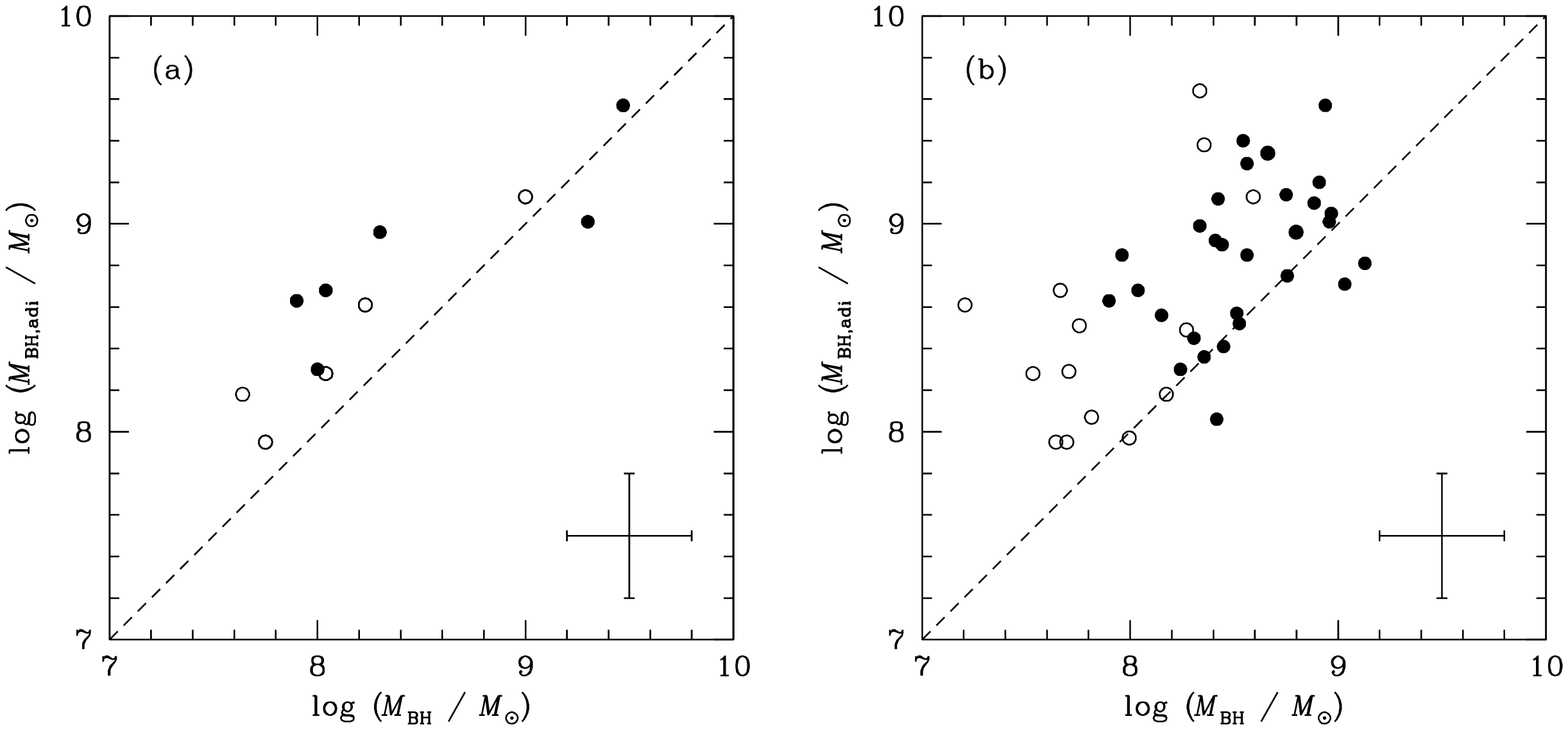,width=19.5cm,angle=90}}
\figcaption[f3.ps]{
Comparison of BH masses obtained from the adiabatic-growth models of
van~der~Marel (1999a) with ({\it a}) BH masses derived directly using stellar
and gas kinematics (Kormendy \& Gebhardt 2001) and ({\it b}) BH masses derived
using the $M_{\rm BH}-\sigma$ relation from Gebhardt et al. (2000).  The solid
points correspond to core galaxies and open circles to power-law galaxies.
Representative error bars are given in the lower-right corner of the plot.
\label{fig3}}
\end{figure*}
\vskip 0.3cm


\noindent
correlation 
coefficient is 0.57, with a probability of $5 \times 10^{-4}$ that the two 
variables are uncorrelated.  However, the observed slope is substantially 
steeper than the unity slope suggested by Nakano \& Makino (1999).  The 
best-fitting linear regression line, obtained by adopting $M_{\rm BH}$ as the 
independent variable and including nominal errors on both variables, has a 
slope of $\sim 2.8\pm 0.84$.  It would be fruitful in the future to 
reexamine this issue using simulations that incorporate more realistic 
initial conditions.
 
According to the merger models, more luminous galaxies should have experienced
a larger number of mergers than fainter galaxies, and hence their centers
should have been ``scoured'' more thoroughly by the orbital decay of BH 
binaries.  This scenario provides an attractive explanation for the observed 
correlation between core size and total bulge luminosity (Faber et al. 1997; 
Ravindranath et al. 2001b).  Recently, Milosavljevi\'c \& Merritt (2001) 
carried out more realistic simulations that start with galaxies containing 
central 
BHs and steep density cusps ($\rho \propto r^{-2}$), as observed in power-law 
galaxies. Unlike earlier simulations where the dynamical effects of the massive 
BH were studied by introducing a naked BH into an existing core, their model 
includes the effect of stars bound to the BHs. This is found to reduce the 
timescale over which the two BHs approach each other due to dynamical friction 
and form a binary. As the binary orbit shrinks, the gravitational slingshot 
effect ejects stars from the center and causes the density cusp to flatten to 
$\rho \sim r^{- \gamma^{\prime}}$, where $\gamma^{\prime} < 2$. Assuming that 
the initial central density profiles are represented by steep 
$\rho \sim r^{-2}$ cusps, the ejected mass or the mass deficit is defined as 
the mass needed to recover the initial profile from the observed flat profile 
with slope $\gamma^{\prime}$. A key 
prediction from the simulations by Milosavljevi\'c \& Merritt (2001) is that 
the mass ejected during each merger event by the BH binary should be 
comparable to their combined mass.  The total mass ejected depends on the 
final BH mass and the number of mergers, and hence a steeper-than-linear 
relation is expected between the mass deficit and BH mass.  Milosavljevi\'c \& 
Merritt (2001) give the following expression for the ejected mass:

\begin{equation}
M_{\rm ej} \approx \frac{2(2-\gamma^{\prime})}{3-\gamma^{\prime}} 
\frac{\sigma^{2}r_{b}}{G},
\end{equation}

\noindent
where $r_b$ is the core radius (equivalent to the break radius in the Nuker
function) and $\gamma^{\prime}$ is the inner-cusp slope of the luminosity 
density profile.   We obtained the luminosity density profile by inverting the 
surface brightness profile, parameterized by the Nuker function, using the 
Abel integral equation and assuming spherical symmetry (Binney \& Tremaine 
1987).  A plot of $M_{\rm ej}$ versus $M_{\rm BH}$ (Fig.~4{\it b}) shows that 
the two quantities are indeed well correlated.  The best-fitting linear 
regression line, again obtained by adopting $M_{\rm BH}$ as the independent 
variable and including nominal errors on both variables, gives a slope of 
$\sim 2.3\pm 0.51$.  Milosavljevi\'c \& Merritt (2001) presented a similar 
result based on a smaller sample. Formally, the correlation in our sample 
again is statistically very significant; the Spearman's $\rho$ correlation 
coefficient is 0.77, and the probability that the correlation could have 
arisen by chance is $< 10^{-4}$.

We stress that the tightness of the $M_{\rm BH}-\sigma$ relation
(Gebhardt et al. 2000; Ferrarese \& Merritt 2000) implies that there should
be little difference between BH masses estimated in this fashion compared to
directly measured masses.  We illustrate this point in Figure~4 by replotting
in open symbols the six objects having kinematically determined BH masses.

\section{Discussion and Summary}          

As discussed by Faber et al. (1997) and Ravindranath et al. (2001b), the cores 
of luminous elliptical galaxies populate a core ``fundamental plane'' (defined 
by $r_{b}$, surface brightness $\mu_{b}$ at $r_{b}$, and $\sigma_0$), 
analogous to the fundamental plane on global scales (defined by $r_{e}$, the 
surface brightness $\mu_{e}$ at $r_{e}$, and $\sigma_e$).  More luminous 
galaxies with higher central velocity dispersions have characteristically 
larger, more diffuse cores.  Most 
%
\begin{figure*}[t]
\figurenum{4}
\centerline{\psfig{file=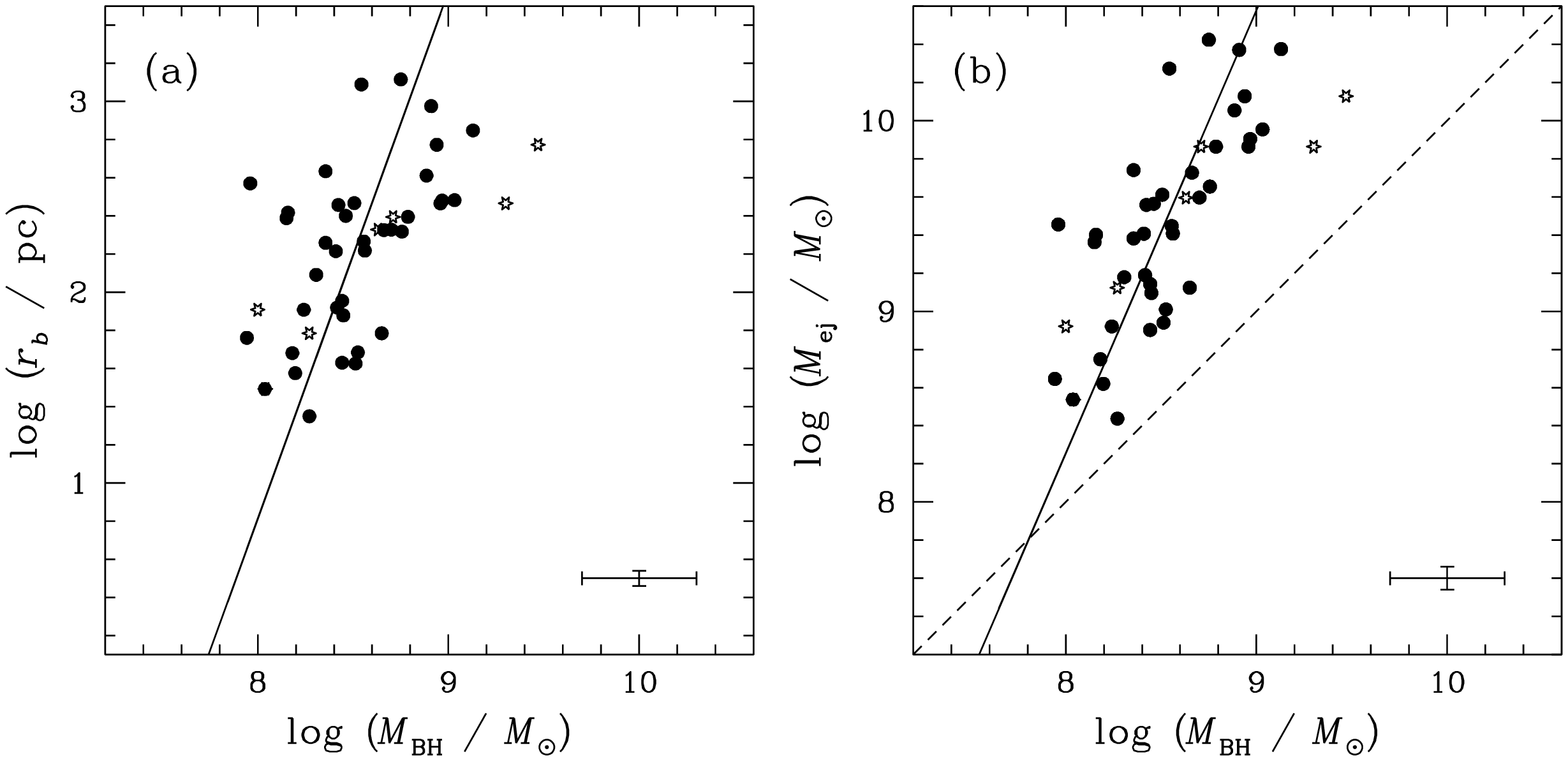,width=19.5cm,angle=90}}
\figcaption[f4.ps]{
The relation among core galaxies between BH mass $M_{\rm BH}$ and ({\it a}) 
the break radius $r_b$ and ({\it b}) the ejected mass $M_{\rm ej}$. The solid 
line in each panel gives the best-fitting linear regression (see text).  The 
dashed line in panel ({\it b}) denotes $M_{\rm ej} = M_{\rm BH}$. The open 
star symbols replot the six galaxies whose BH masses have been measured 
directly by kinematics.  They do not show any obvious systematic deviations 
compared to the solid points. Representative error bars are given in the 
lower-right corner of the plot.
\label{fig4}}
\end{figure*}
\vskip 0.3cm

\noindent
theories for the formation of early-type
galaxies involve dissipative gas inflows that tend to increase the central
density and form steep nuclear cusps. While low-luminosity galaxies with
power-law profiles can be explained naturally in this manner, the
existence of diffuse cores with weak cusps in massive galaxies poses a serious
observational challenge, as discussed by Faber et al. (1997).  How did they
form?  And once created, how did they manage to avoid acquiring high-density
cores from the accretion of low-mass satellites?

A promising framework to address the formation of cores invokes the dynamical
effects of merging central massive BHs.  This is all the more appealing
because there is now increasing evidence that BHs are ubiquitous in bulges
(Magorrian et al. 1998; Ho 1999; Kormendy \& Gebhardt 2001).  If so, the 
hierarchical assembly of massive elliptical galaxies would inevitably lead to 
BH mergers.  The most recent simulations by Milosavljevi\'c \& Merritt 
(2001) incorporate realistic initial conditions for the merging galaxies. 
Starting with two low-mass galaxies, each containing a central BH and a steep 
nuclear cusp, the hardening of the BH binary leads to the formation of a 
remnant with a large, diffuse core.  We show in this paper that galaxy cores 
obey certain trends that are expected for the merger remnants of binary BHs.

If cores form as an aftermath of galaxy mergers containing BHs, the work of 
Nakano \& Makino (1999) indicates that the sizes of the cores should scale 
with the final BH mass.  Using the $M_{\rm BH}-\sigma$ relation to estimate BH 
mass, we find that this indeed seems to hold in our sample: the core size 
increases roughly as $r_{b} \propto M_{\rm BH}^{2.8}$ (Fig.~4{\it a}).  Although 
the slope of the correlation is steeper than predicted by Nakano \& Makino 
(1999), the discrepancy should not be taken too literally given the simplified 
initial conditions used in their numerical calculations.  Combining this 
functional dependence with $M_{\rm BH} \propto \sigma^{3.8}$ (Gebhardt et al. 
2000), we recover within the errors, the empirical scaling relation between 
core size and velocity dispersion, $r_{b} \propto \sigma^6$ (Faber et al. 1997;
Ravindranath et al. 2001b).  Thus, in this interpretation, the scaling relation
observed between $r_{b}$ and $\sigma$ is an indirect manifestation of two more 
fundamental correlations, namely the $r_{b} - M_{\rm BH}$ relation which 
reflects the dynamical effects of binary BHs during core formation, 
and the $M_{\rm BH} - \sigma$ relation which bears on processes associated 
with the joint formation of BHs and bulges (e.g., Kauffmann \& Haehnelt 2000; 
Burkert \& Silk 2001).

Using the numerical models of Quinlan \& Hernquist (1997), Faber et al. (1997)
already demonstrated that BH mergers produce cores whose masses scale 
approximately with the final BH mass.  They suggested that the 
relation between core and BH mass might depend on the slope of the nuclear 
cusp and pointed out the need for simulations with more realistic initial 
conditions.   The calculations of Milosavljevi\'c \& Merritt (2001) add 
a new degree of realism to the simulations by modeling the 
merging galaxies as systems with initially steep cusps, as seen in 
power-law galaxies.  Adopting their formalism to compute the amount of 
stellar mass ejected by the BH binary during core formation, and again the 
$M_{\rm BH}-\sigma$ relation to estimate BH mass, we find that 
$M_{\rm ej} \propto M^{2.3}_{\rm BH}$ (Fig.~4{\it b}).  Unlike the case of the 
relation between $r_b$ and $M_{\rm BH}$, the slope of the correlation between 
$M_{\rm ej}$ and $M_{\rm BH}$ cannot be easily traced to known 
central-parameter scaling relations.  The $r_b - \sigma$ and 
$M_{\rm BH}-\sigma$ relations, when combined with Equation 2, yields a 
slope of $\sim$4 for the $M_{\rm ej}-M_{\rm BH}$ relation, whereas the 
actual observed slope is $\sim$2.  Although each merger produces 
$M_{\rm ej} \approx M_{\rm BH}$, the total ejected mass depends on the 
number of mergers; more massive galaxies with more massive BHs require a 
larger number of mergers, and thus the slope of the $M_{\rm ej} - M_{\rm BH}$ 
is expected to be $> 1$.  This is in qualitative agreement with the 
observations.  In detail, of course, the exact slope should depend on 
complications not treated in the models.  These include heterogeneous 
mergers and additional processes such as gas accretion.

A competing view of core formation posits that the nuclear cusps imprint the 
formation of the central BH through adiabatic growth. According to this 
picture, the slope of the cusp profile should correlate with the relative BH 
mass fraction; for a galaxy core of a given mass, a more massive BH induces a 
steeper cusp. This basic expectation, however, is not supported by the data. 
The strongest disagreement comes from the power-law galaxies, whose steep 
cusps would suggest that these objects have exceptionally large BH mass 
fractions, contrary to what is actually observed.  The core galaxies fare 
somewhat better. The distribution of cusp slopes versus BH mass fraction is 
not inconsistent with the sharp rise anticipated by the models (Fig.~2), 
although the agreement cannot be said to be good.  Regardless of this 
comparison, however, a basic shortcoming of the adiabatic-growth model is
that it leaves the main problem of the formation of cores unanswered; the 
BH is assumed to grow in a preexisting core.

As shown by van~der~Marel (1999a), one of the most attractive features of the
adiabatic model is that it potentially allows us to estimate BH masses based on
photometric data alone. This would be a remarkably efficient alternative to the
time-consuming spectroscopic observations needed for the conventional methods
based on kinematics.  In view of the difficulties described above, it might 
come as a surprise that the BH masses derived from the adiabatic models should
differ from the actual masses by just $\sim$0.5 dex (Fig.~3). This is not
entirely unexpected, however, given that the adiabatic models were fitted
specifically to the core region of the central light profiles, whose size,
$r_b$ (see Table~1A), is typically only a factor of a few larger than the
``sphere of influence'' of the BH, $r_{\rm BH}\,\simeq\,G M_{\rm BH}/
\sigma^2$.  In other words, the mass scale of the initial core, set by the core
size, is comparable to the BH mass.  In \S~3.1, we noted that the offset
between the model-inferred masses and the actual masses tends to be
systematically larger for power-law galaxies than it is for core galaxies
(Fig.~3).  This is a simple consequence of the fact that the adiabatic models
need to invoke larger BH mass fractions ($\mu$) to fit power-law galaxies than
in core galaxies, whereas empirically both types of galaxies share rather
similar BH mass fractions. 

In summary, we find that the existing data on the central regions of early-type 
galaxies support the notion that the observed properties of the cores in 
luminous galaxies have been substantially molded by the dynamical influence 
of binary massive BHs during past merger events.  The merger hypothesis can 
consistently account for the core fundamental-plane relations and the 
dependence of core size and mass on central BH mass. The observed cusp 
slopes do not appear to comply with the expectations of the adiabatic-growth 
model for BHs.

\acknowledgements
This work is funded by NASA LTSA grant NAG 5-3556, and by NASA/{\it HST}\ grants
AR-07527 and AR-08361 from the Space Telescope Science Institute (operated by 
AURA, Inc., under NASA contract NAS5-26555). A.~V.~F. is grateful to the 
Guggenheim Foundation for a Fellowship. We thank Chien Y. Peng for writing the 
program to compute the luminosity density profiles, Roeland P. van~der~Marel 
for helpful discussions on the adiabatic-growth models, and the referee 
for constructive criticisms.  We made use of the NASA/IPAC Extragalactic 
Database (NED) which is operated by the Jet Propulsion Laboratory, California 
Institute of Technology, under contract with NASA, and of the Lyon-Meudon 
Extragalactic Database.


\end{document}